# Investigation of the presence of charge order in magnetite by measurement of the spin wave spectrum

R. J. McQueeney,[1,*] M. Yethiraj,[2] W. Montfrooij,[3] J. S. Gardner,[4,5]

P. Metcalf,[6] J. Honig[6]

[1]*Department of Physics & Astronomy and Ames Laboratory, Iowa State University, Ames, IA 50011 USA*

[2]*Center for Neutron Scattering, Oak Ridge National Laboratory, Oak Ridge, TN 37831 USA*

[3]*Department of Physics and Missouri Research Reactor, University of Missouri, Columbia, MO 65211 USA*

[4]*Physics Department, Brookhaven National Laboratory, Upton, NY 11973 USA*

[5]*NIST Center for Neutron Research, National Institutes of Standards and Technology, Gaithersburg, MD 20899 USA*

[6]*Department of Chemistry, Purdue University, West Lafayette, IN 47907 USA*

(Dated: February 7, 2006)

## ABSTRACT

Inelastic neutron scattering results on magnetite ($Fe_3O_4$) show a large splitting in the acoustic spin wave branch, producing a 7 meV gap midway to the Brillouin zone boundary at **q** = (0,0,1/2) and $\hbar\omega$ = 43 meV. The splitting occurs below the Verwey transition temperature, where a metal-insulator transition occurs simultaneously with a structural transformation, supposedly caused by the charge ordering on the iron sublattice. The wavevector (0,0,1/2) corresponds to the superlattice peak in the low symmetry structure. The dependence of the magnetic superexchange on changes in the crystal structure and ionic configurations that occur below the Verwey transition affect the spin wave dispersion. To better understand the origin of the observed splitting, we have constructed a series of Heisenberg models intended to reproduce the pairwise variation of the magnetic superexchange arising from both small crystalline distortions and charge ordering. We find that none of the models studied predicts the observed splitting, whose origin may arise from charge-density wave formation or magnetoelastic coupling.

# I. INTRODUCTION

Magnetite (Fe$_3$O$_4$) was the first magnetic material to ever be discovered and utilized. It still has widespread usage in modern society as it is a rather strong, permanent ferrimagnet ($T_c$ = 858 K). While its magnetic properties are well known, magnetite surprisingly remains one of the more controversial examples of a correlated electron system. In 1939, Verwey discovered that magnetite undergoes a metal-to-insulator transition, resulting in a decrease of the conductivity by two orders-of-magnitude below the Verwey temperature, $T_V$ = 123 K.[1] Verwey originally postulated that the hopping of extra electrons residing on the spinel B-site iron sublattice is responsible for the metallic conductivity. In the insulating phase, these extra electrons freeze out in an ordered pattern due to their mutual Coulombic repulsions.[2] This charge ordering transition, called a Verwey transition, is one of the earliest instances of invoking many-body effects to explain a solid-state phase transition. In the intervening years, Verwey's hypothesis has survived in some shape or form. Neutron diffraction eventually demonstrated that the symmetry lowering predicted by Verwey's original charge ordering model cannot be entirely correct.[3] Several other charge ordering schemes have been proposed that are consistent with the low temperature monoclinic crystal symmetry.[4,5] Due to the complexity of the low temperature structure (224 atoms/unit cell) and twinning-related multiple monoclinic domains in the low temperature phase, conclusive evidence for the existence of charge ordering is a difficult claim. So difficult, in fact, that new evidence is being put forth that raises serious doubts that magnetite is charge ordered at all at low temperatures,[6] opening a new dialogue about this venerable system.[7,8]

One property that should be very sensitive to the presence of charge ordering is the spin wave spectrum. At room temperature, the spin wave dispersion is consistent with a simple nearest-neighbor Heisenberg Hamiltonian. The magnetic superexchange interactions between iron pairs is mediated by oxygen anions and have been determined by inelastic neutron scattering.[9,10] If charge ordering is present below $T_V$, the superexchange values will be modified in a periodic way due to the charge (valence) ordering. This modification will appear as a splitting of the spin wave spectrum at wavevectors corresponding to the charge ordering wavevector.

In this paper, we report inelastic neutron scattering results that clearly show a large (7 meV) splitting in the acoustic spin wave branch below $T_V$. The splitting occurs midway to the Brillouin zone boundary at $\mathbf{q} = (0,0,1/2)$ and $\hbar\omega = 43$ meV. The wavevector $(0,0,1/2)$ corresponds to the cell doubling supposedly caused by charge ordering. To better understand the origin of the splitting, we have constructed a series of Heisenberg models intended to reproduce the pairwise variation of the magnetic superexchange arising from both small crystalline distortions and charge ordering. We find that none of the models studied predicts the observed splitting. A preliminary report of the results of these measurements has been published elsewhere.[11]

## II. EXPERIMENTAL

### A. Samples

Above $T_V$, magnetite adopts a cubic inverse spinel crystal structure with $Fd\bar{3}m$ space group and a lattice constant of $a_c = 8.394$ Å. The primitive rhombohedral unit cell consists of six iron atoms and eight oxygen atoms whose positions are given in

Table I. There are two different symmetry iron sites, labeled A and B. The two A-sites reside in tetrahedrally coordinated oxygen interstices, the four B-sites have octahedral coordination. From valence counting, the A-site is $3^+$ ($3d^5$ electron configuration), and the B-site has a fractional average valence of $2.5^+$. The magnetic structure is that of a collinear ferrimagnet with A and B moments aligned antiparallel.

Below 123 K, magnetite undergoes a first-order metal-insulator transition. The transition lowers the crystallographic symmetry from cubic to the monoclinic $Cc$ space group. The monoclinic structure consists of slight distortions from a superstructure of the cubic cell of dimensions $\sqrt{2}a_c \times \sqrt{2}a_c \times 2a_c$ ($a_m$ = 11.868 Å, $b_m$=11.851 Å, $c_m$=16.752 Å, $\beta$=90.2°). There are 32 formula units in the $Cc$ cell and a table of the atomic positions can be found in Ref. 4. Ignoring the small monoclinic distortion, the low temperature structure can also be described in a slightly less complicated orthorhombic structure in the $Pmca$ space group. This is related to a cubic supercell of size $a_c/\sqrt{2} \times a_c/\sqrt{2} \times 2a_c$. There are 8 formula units in the $Pmca$ cell and the atomic positions are listed in Ref. 12. There are no reports to indicate that the magnetic moments do not remain in a collinear configuration below $T_V$. In discussion of the low temperature phases, we will refer to either the $Cc$ or $Pmca$ cell as the need warrants. As the atomic distortions in the Verwey structure are very small, throughout the paper we will only refer to crystal directions in terms of the ($h,k,l$) indices of the original cubic cell, to avoid confusion.

We used various single-crystal samples weighing from 5 – 10 grams each that were prepared from powdered material by use of a radio frequency induction melting technique.[13] This method permits melting of $Fe_3O_4$ in a crucible lined with solid of the same composition, so as to minimize accidental contamination of the melt. After slow cooling the boule was fractured and single crystals were extracted These crystals were then reannealed in specialized equipment[14] under appropriate $CO_2/CO$ atmospheres so as to achieve the desired oxygen/iron ratio. The Verwey transition temperature for these samples is ~123 K with a narrow transition width, a good indication that the samples are homogeneous, with few metallic impurities, and nearly ideal oxygen stoichiometry.[15] In addition, ideal oxygen stoichiometry is indicated by the observation of a 0.3° splitting due to monoclinic domains.[15] Crystal mosaics were ~0.1° as obtained from neutron diffraction rocking curves.

### B. Triple-axis measurements

We performed triple-axis neutron scattering measurements of the magnetic excitations using several different instruments under a variety of experimental configurations. Triple axis measurements were made on the HB-3 and HB-1 spectrometers at the High Flux Isotope Reactor at Oak Ridge National Laboratory and the C-5 spectrometer at the NRU at Chalk River Laboratories. Detailed experimental configurations are shown in Table II and will be referred to in the text by the configuration label. Measurements were made in horizontal magnetic field as well as zero applied magnetic field. The horizontal field is applied along the cubic [001] direction. The applied magnetic field served two purposes: 1) a single magnetic domain sample can be created in a modest magnetic field of $H < 1$ Tesla, 2) in the low temperature phase, field-cooling with $H \sim 1$ Tesla will cause the cell doubling direction (the *c*-axis of the monoclinic structure) to orient along the applied field direction.[3] Each of the three equivalent cubic axes become the *c*-axis of the monoclinic phase. Hence, there are two monoclinic domains for each cubic domain. However, the two domains are only separated by 0.3° and are treated as a single domain for the inelastic scattering measurements.

### C. Observation of the gap

Above the Verwey transition, the spin wave dispersion can be well understood as a simple Heisenberg ferrimagnet. The dispersion consists of six branches, one acoustic and five optical. Brockhouse originally measured the acoustic branch and one optical branch[9] and these dispersion data were fit to a Heisenberg model with nearest-neighbor superexchange parameters $J_{AB}$ = -2.4 meV and $J_{BB}$ = 0.24 meV.[10] We have performed measurements of the acoustic dispersion along the [001] direction which are consistent with Brockhouse's original work above $T_V$. This will be discussed in more detail below. At temperatures below the Verwey transition, we observe clear evidence of the splitting of the acoustic branch at **q** = (0,0,1/2) (in units of $2\pi/a_c$ referencing the cubic cell). We define the reduced wavevector **q** as that part of the momentum transfer (scattering) vector $\hbar\mathbf{Q}$ to lie within the first Brillouin zone such that $\mathbf{Q} = \boldsymbol{\tau} + \mathbf{q}$, where $\boldsymbol{\tau}$ is a reciprocal lattice vector of the cubic structure. Brockhouse and co-authors had previously studied the dispersion below $T_V$, but reported no substantive change compared to the room temperature dispersion.[15] We assume this negative result was due to the use of natural crystals where impurities can have deleterious effects on the Verwey transition.[7] Using high-quality synthetic single-crystals, we observe a rather large splitting of the acoustic branch below $T_V$.

Clearest indication of the nature of the gap is seen in a scattering configuration with a horizontal magnetic field of 1.1 Tesla applied along the cubic [001] direction. Measurements were made on the C-5 spectrometer at Chalk River Laboratories in configuration A of Table II. Figure 1 shows the orientation of the magnetic field in the [h0l] scattering plane, typical measured momentum transfer vectors, and the Brillouin zone boundaries of both the cubic ($T > T_V$) and orthorhombic ($T < T_V$) crystal structures. Figure 2 shows constant-**Q** energy scans with momentum transfers **Q** = (0,0,4.5), (4.5,0,0), (4,0,1/2), and (-1/2,0,4) each corresponding to cubically-equivalent (0,0,1/2)-type reduced wavevectors with an acoustic spin wave energy of $\hbar\omega$ = 43 meV. In the orthorhombic phase, the **q** = (0,0,1/2) and (1/2,0,0) reduced wavevectors are inequivalent, as is clear from Fig. 1, with (0,0,1/2) being a Brillouin zone center of the orthorhombic structure. The acoustic spin wave excitation is clearly split below $T_V$ into two distinct modes at 39 and 46 meV for **q** = (0,0,1/2) (Figs. 2(a) and 2(b)). Weak evidence for splitting can also be seen for **q** = (1/2,0,0) (Figs. 2(c) and 2(d)). The main splitting therefore occurs only when **q** is along the cell doubling direction of the orthorhombic structure. The small residual splitting remaining at (1/2,0,0) wavevectors may be due to incomplete removal of domains causing a minority of cell doubling domains to occur along the [100] direction. It is also clear that the two excitation peaks below $T_V$ appear narrower than the single peak above $T_V$. We will address the intrinsic widths versus resolution widths for these peaks in Section III below.

Figure 2 also shows that the spin wave intensity is approximately twice as large for **Q** along [001] (Figs. 2(b) and 2(c)) than along [100] (Figs. 2(a) and 2(d)). This is due to the vector nature of the neutron spin scattering cross-section. The scattered intensity for a ferrimagnet aligned in a magnetic field along the z-direction is,

$$S(\mathbf{Q},\omega) \propto \left(1 + \frac{Q_z^2}{|\mathbf{Q}|^2}\right). \tag{1}$$

The momentum transfer vector lying along the field direction will have twice the signal of a scattering vector perpendicular to the field, as is clear from Fig. 2.

The spin wave splitting occurs only below $T_V$. Figure 3(b) shows the temperature dependence of the intensity in the gap at $\mathbf{q} = (0,0,1/2)$ and $\hbar\omega = 43$ meV. This measurement was performed using configuration B without a horizontal magnetic field. Below $T_V$, the sample contains all three orthorhombic c-axis domains, hence the ratio of intensities measured at the gap energy above and below $T_V$ is not as large as Fig. 2(a). The gap begins to form below the Verwey temperature, and can be compared to the order parameter of the (6,0,1/2) superlattice peak in Fig. 3(a).

### D. Extent of the gap in q-space

Figure 4 displays constant-**Q** energy scans along the [001] direction as measured in configuration C. Figure 4(a) shows a set of scans above the Verwey transition at room temperature and Fig. 4(b) shows scans along the same direction at $T = 115$ K. As configuration C has no applied magnetic field, the resulting spin waves observed along [001] at $T = 115$ K are actually averaged over all three orthorhombic principal directions. Similar to Fig. 3(b), the gap is still observed despite the domain averaging. In each figure, the peaks of the spin wave excitations are marked with a gray circle. The spin wave gap is indicated by the red slanted line and appears to have some **q**-dependence. The dotted line indicates the gap energy of 43 meV at (0,0,4.5). Phonon excitations were also observed and are indicated by the dashed lines. Precise fits to the low temperature dispersion will be discussed in Section III, as the resolution effects are important in the interpretation of the data.

The HB1 and HB3 spectrometers were used to study the extent of the splitting in other directions in reciprocal space. Using the B and D configurations, a series of constant energy scans were measured on the [h,0,4+l] plane above and below the Verwey transition. Mesh scans were performed at $\hbar\omega$ = 39 meV, 43 meV, and 46 meV, the energy positions of the lower peak, the gap, and the upper peak of the split mode at (0,0,1/2). Figure 5 shows false color contour plot of the scattered intensity and Fig. 6 shows line plots of the constant energy cuts along the (h,0,4.5) direction. Above the transition, the contour plots demonstrate that the spin wave dispersion is isotropic and forms a circular ring of intensity around the (004) Brillouin zone center. In the Verwey phase, the gap at 43 meV is clearly seen and extends ± 0.1 – 0.2 rlu in the transverse direction along [1,0,0]. At 39 meV, it does not appear that the spin waves are affected to any great extent. At 46 meV, a suppression of the intensity is observed near the (0.25,0,4.5) position at low temperatures as shown in Figs. 5 and 6. This is feature is not another gap, but is rather the original gap at (0,0,4.5) picked up by the experimental resolution of the instrument. We will delay further discussion of important resolution effects until the next section, where we consider the effect of experimental resolution on interpretation of the gap structure.

## III. DATA ANALYSIS AND MODELLING

Before making detailed analyses of the dispersion data, we introduce model calculations for the spin wave scattering cross-section for magnetite in the cubic spinel phase. This model can be used to perform resolution convolutions of the spin wave cross-section for comparison to the data. Once the machinery is established, its use can be extended for the more complicated low symmetry phase occurring below $T_V$. We start with a discussion of the local ionic states and build up to the collective spin excitations.

### A. Magnetic states and excitation spectrum of magnetite

Following primarily the work of Buyers et al.,[17] the general Hamiltonian for the lowest energy magnetic states of the iron atoms in magnetite can be written

$$H = H_{Hund} + H_{CF} + H_{SO} + H_{ex} \tag{2}$$

where $H_{Hund}$ describes the intra-atomic electronic correlations, $H_{CF}$ the crystalline electric field, $H_{SO}$ the spin-orbit coupling in the Fe *3d* orbitals. $H_{ex}$ is the exchange interaction between atoms. The Hamiltonian can be further categorized as containing two terms $H=H^{(1)}+H^{(2)}$. $H^{(1)}$ is a single-ion term

$$H^{(1)} = H_{Hund} + H_{CF} + H_{SO} + H_{mf} \tag{3}$$

where $H_{mf}$ describes the molecular field arising from ferrimagnetic long-range order. The residual term containing exchange interactions between the ions, $H^{(2)}$, will describe the spin waves in the ordered state.

$$H^{(2)} = H_{ex} - H_{mf} \tag{4}$$

**B. Single-ion term**

In magnetite, the strength of each term in $H^{(1)}$ has the following order: $H_{Hund} \gg H_{CF} \gg (H_{SO}, H_{mf})$. In this weak-field limit, we consider only the ground state term of the Hund multiplet, with higher terms being several eV above the Hund's rule ground state. The crystal field is the next strongest term in the Hamiltonian and will split the ground state Hund's rule term. The two remaining interactions are generally weaker than the crystal field and will determine the details of the ionic configuration.

In the high temperature cubic inverse spinel structure of magnetite, there are two crystallographically distinct sites; the tetrahedral A-site and the octahedral B-site. The A-sites contain only $Fe^{3+}$ ions with a $3d^5$ electronic configuration, giving a singlet $^6S_{5/2}$ Hund's ground state term. The point group symmetry of the A-site is cubic ($T_d$), however this crystalline electric field cannot not split the orbital singlet ground state. The ground state also has no spin-orbit splitting, since the orbital moment is zero. Therefore, $Fe^{3+}$ has a spin-only ground state S = 5/2.

The B-sites contain both $Fe^{3+}$ and $Fe^{2+}$ states. Similar to the A-site, the B-site $Fe^{3+}$ Hund's rule singlet ground state is unsplit by crystal field and spin-orbit interactions and has a S = 5/2 ground state. On the other hand, the $Fe^{2+}$ ion has a $^5D_4$ ground state that will be split by the crystal field and spin-orbit interactions. The B-site actually possesses a trigonal ($D_{3d}$) point group symmetry in the cubic $Fd\bar{3}m$ space group ($H_{CF} = H_{cubic} + H_{trigonal}$). The trigonal component of this field ($H_{trigonal} \sim 150$ meV), which is due to neighbors beyond the first shell, is weak in comparison to the nearest-neighbor oxygen octahedral field ($H_{cubic} \sim 1.5$ eV, $H_{trigonal} \ll H_{cubic}$).[18] The larger octahedral crystal field splits the $^5D_4$ ground state into a triply degenerate ground state ($^5T_2$) and an excited state doublet ($^5E$) with the $^5E$ state sufficiently high in energy that we may disregard it from now on. The weaker trigonal field will further split the cubic $^5T_2$ ground state into a ground state singlet ($^5A_1$) and excited state doublet ($^5E$). For cubic symmetry, with the trigonal axis $\alpha \approx 60°$, only minor mixing occurs between the two $^5E$ states.[19] Spin-orbit interactions are expected to be even smaller than the trigonal field ($E_{SO} \sim 10$ meV),[18,20,21] splitting the $^5E$ orbital doublet into five doubly degenerate $|ls\rangle$ states and leaving the $^5A_1$ orbital singlet ground state unchanged.

Finally, we consider the molecular field acting on the both the $Fe^{2+}$ and $Fe^{3+}$ ions. At room temperature and below, we can assume that the magnetization is nearly saturated in the ferrimagnetic state with $T_C = 858$ K. The molecular field at low temperatures can then be estimated from the Curie constants on each sublattice, the Curie temperature ($T_C$), and the saturation magnetization ($M_s$). The molecular field is

$$B_{mf} = \frac{T_c}{\sqrt{C_A C_B}} M_s \qquad (5)$$

where the Curie constants are

$$C_i = \frac{N_i g_i^2 J_i (J_i + 1) \mu_B^2}{3 k_B} . \qquad (6)$$

For magnetite with $M_s = 510$ G, $T_C = 858$ K, $C_A = 0.1$ K ($J_A = 5/2$, $g_A = 2$), and $C_B \approx 0.16$ K ($J_B \approx 2.25$, $g_B \approx 2$) gives a molecular field of $\sim 350$ T. For the $Fe^{3+}$ ($Fe^{2+}$) ions, the maximum Zeeman splitting in the molecular field is $gS\mu_B B_{mf} \approx 100$ meV (80 meV). Figure 7 shows a schematic drawing of the spectrum of single-ion states in cubic magnetite.

The above treatment of the local electronic states is strictly for an insulator. We have ignored that the cubic phase is poorly metallic, with electronic bandwidths of order 1 eV.[22,23] Therefore, the electronic hybridization will mix the trigonal crystal field ground states, thereby returning some orbital degeneracy to the B-site iron atoms.

Below $T_V$, the monoclinic distortion will lower the the point group symmetry of each iron site to either $1$, $\bar{1}$, or $m$ and open a gap in the electronic density-of-states. In the cubic spinel phase, the local ground state electronic configurations for $Fe^{2+}$ or $Fe^{3+}$ ions are all orbital singlets. This symmetry lowering cannot split any of the ground state ionic configurations, but it will split the $^5E$ excited state by a small amount (of order 10 meV). Therefore, we do not expect the monoclinic distortion to produce any additional low energy crystal field excitations (< ~ 150 meV) on any of the iron sites.

Using these arguments, we can assume that there are no low energy (< ~ 100 meV) crystalline electric field or spin-orbit excitations either in the cubic or monoclinic phase. These local states have no orbital freedom and can be treated as spin variables in a Heisenberg treatment of the collective excitations.

### C. Collective spin wave excitations

Starting from the local orbitally-quenched ionic configurations, we can now study the low energy collective spin wave excitations given by $H^{(2)}$. A general Heisenberg Hamiltonian is chosen to represent the interactions between the local moments. In the absence of an applied magnetic field, the Heisenberg Hamiltonian is given by

$$H^{(2)} = -\sum_{<li,l'j>} J(li;l'j) \mathbf{S}_{li} \cdot \mathbf{S}_{l'j} \qquad (7)$$

where $J(li;l'j)$ are the pairwise exchange values between ground state configurations of the $i$th atom in the $l$th unit cell and the $j$th atom in the $l'$th unit cell. Single-ion anisotropy terms are extremely small in magnetite (with anisotropy fields of 0.1-1 T, depending on temperature) and are set to zero.[10,24,25] For an arbitrary number of collinear spins in the unit cell, Saenz[26] has developed a formalism to calculate the spin wave excitation energies, eigenvectors, and neutron scattering intensities. $\mathbf{S}_{li}$ is the vector spin operator for the ground state ion with spin magnitude $\sigma_i S_i$ where $S_i$ is positive and $\sigma_i = \pm 1$ with +1(-1) parallel (antiparallel) to the $z$-quantization axis. For the ferrimagnetic magnetite structure, $\sigma_A = -1$ and $\sigma_B = +1$. After performing the Holstein-Primakoff transformation, the secular equation for the system is $(\underline{\mathbf{M}}-\lambda\underline{\mathbf{I}})\mathbf{T} = 0$ where

$$M_{ij}(\mathbf{q}) = 2\delta_{ij}\sum_{lk} J(0i;lk)\sigma_k S_k - 2\sigma_j\sqrt{S_i S_j}\sum_l J(0i;lj)exp(i\mathbf{q}\cdot\mathbf{l}). \tag{8}$$

The eigenvalues, $\gamma_n(\mathbf{q})$, and eigenvectors, $T_n(\mathbf{q})$, are obtained by diagonalization of the secular matrix $\underline{\mathbf{M}}(\mathbf{q})$ at wavevector $\mathbf{q}$, where $n$ labels the spin wave branch. Inspection of the matrix shows that it is not Hermitian, due to $\sigma_j$. However, the eigenvalues for this matrix are purely real and an entire branch must either be entirely positive or entirely negative (i.e. no branch ever crosses zero), with the number of negative branches equal to the number of antiparallel spin sites. Thus, the spin wave dispersion for branch $n$ is $\hbar\omega_n(\mathbf{q}) = |\gamma_n(\mathbf{q})|$. The normalized eigenvector has components

$$T_{ni} = \sqrt{S_i}\xi_{ni} / \sqrt{\sum_i S_i\xi_{ni}^2} \tag{9}$$

where $\xi_{ni}^2$ is the fraction of the $i$th spin contained in the eigenvector and $\sum_i T_{ni}^2 = 1$ for each branch. This definition of the eigenvector is used to calculate the neutron cross-section.

## D. Neutron scattering cross-section for spin waves

The neutron cross-section for unpolarized magnetic scattering is:

$$\frac{d^2\sigma}{d\Omega dE'} = r_0^2 \frac{k_f}{k_i} S(\mathbf{Q},\omega) \tag{10}$$

where $k_i$ and $k_f$ are the initial and final neutron wavenumbers, $r_0$ is the classical electron radius, and $S(\mathbf{Q},\omega)$ is the Fourier transform of the spin-spin correlation function

$$S(\mathbf{Q},\omega) = \frac{1}{2\pi\hbar}\sum_{\alpha\beta}(\delta_{\alpha\beta} - \hat{Q}_\alpha \hat{Q}_\beta)\int_{-\infty}^{\infty} dt\, e^{-i\omega t}\langle \hat{S}_\mathbf{Q}^\alpha \hat{S}_{-\mathbf{Q}}^\beta(t)\rangle \tag{11}$$

In the Heisenberg model, the spin amplitudes are written as the Fourier transform of the spin density

$$\hat{S}_\mathbf{Q}^\alpha(t) = \sum_{li} f_i(Q) e^{-i\mathbf{Q}\cdot(\mathbf{l}+\mathbf{d}_i)} \hat{S}_{li}^\alpha \tag{12}$$

for magnetic ions at sites $\mathbf{d}_i$. The amplitude prefactor $f_i(Q)$ depends on the Lande $g$-factor, the magnetic form factor, $F_i(Q)$, and the Debye-Waller factor, $W_i(Q)$, for the magnetic ion

$$f_i(Q) = \frac{1}{2} g_i F_i(Q) e^{-W_i(Q)}. \tag{13}$$

By expanding the local spin deviation in terms of plane waves

$$\hat{S}_{li}^\alpha(t) = \frac{1}{N}\sum_\mathbf{q} e^{i\mathbf{q}\cdot\mathbf{l}} \hat{S}_{\mathbf{q},i}^\alpha(t) \tag{14}$$

the above correlation function becomes

$$\langle \hat{S}_\mathbf{Q}^\alpha \hat{S}_{-\mathbf{Q}}^\beta(t)\rangle = \sum_{ij} f_i(Q) f_j(Q) e^{i\mathbf{Q}\cdot(\mathbf{d}_j-\mathbf{d}_i)} \sum_\mathbf{q} \delta(\mathbf{Q}-\mathbf{q}-\boldsymbol{\tau}) \langle \hat{S}_{\mathbf{q},i}^\alpha \hat{S}_{-\mathbf{q},j}^\beta(t)\rangle. \tag{15}$$

The evaluation of the thermal averages of the spin-spin correlation functions for spin wave deviations of the type $S^+S^-$ are performed for each branch labeled $n$

$$\langle S_{\mathbf{q},i}^+ S_{\mathbf{q},j}^-(t)\rangle_n = \frac{\sigma_i \sigma_j}{2N_i}\sqrt{S_i S_j}\, T_{ni} T_{nj}^* \left[n_{\mathbf{q},n} \exp(-i\omega_n(\mathbf{q})t)\right] \tag{16}$$

$$\left\langle S^-_{-\mathbf{q},i} S^+_{-\mathbf{q},j}(t) \right\rangle_n = \frac{\sigma_i \sigma_j}{2N_i} \sqrt{S_i S_j} T_{ni} T^*_{nj} \left[ (n_{-\mathbf{q},n} + 1) \exp(i\omega_n(-\mathbf{q})t) \right] \quad (17)$$

where $n_{\mathbf{q}n}$ is the Bose occupation factor and $T_{ni}$ is the contribution of the $i$th atom to the spin wave eigenvector of branch $n$.

For collinear spins aligned (anti) parallel to the $z$-axis, the spin wave cross-section can then be written ($\mathbf{q} = -\mathbf{q}$) as:

$$S(\mathbf{Q},\omega) = (1 + \hat{Q}_z^2) \sum_{\mathbf{q},n} \left| \sum_i f_i(Q) \sigma_i \sqrt{S_i} T_{ni} e^{-i\mathbf{Q}\cdot\mathbf{d}_i} \right|^2 \delta(\mathbf{Q} - \mathbf{q} - \boldsymbol{\tau}) \times$$
$$\left[ n_{\mathbf{q},n} \delta(\omega + \omega_n(\mathbf{q})) + (n_{\mathbf{q},n} + 1) \delta(\omega - \omega_n(\mathbf{q})) \right] \quad (18)$$

For comparison to the measured intensities, the correlation function above is convoluted with the experimental resolution function.

$$I(\mathbf{Q}_0, \omega_0) = \int S(\mathbf{Q},\omega) R(\mathbf{Q} - \mathbf{Q}_0, \omega - \omega_0) d^3\mathbf{Q} d\omega \quad (19)$$

The resolution function, $R(\mathbf{Q},\omega)$, is calculated from the experimental configuration parameters in Table II and other information, such as the crystal mosaic spreads, using the Cooper-Nathans formalism.[27] Convolutions of Heisenberg model results and other analytical approximations to the dispersion relations were performed with the RESLIB program.[28]

### E. Spin wave calculations for magnetite in cubic spinel phase

At high temperatures, magnetite has the cubic spinel structure in the $Fd\bar{3}m$ space group. The iron atoms in the tetrahedral interstices (A-sites) have a valence of $Fe^{3+}$. Iron atoms in octahedral interstices (B-sites) have an average valence of $Fe^{2.5+}$. The positions and spins of the iron atoms in the cubic primitive cell are given in Table III. Table IV lists the pairwise Heisenberg exchange values used for cubic magnetite.[10] Figure 8 shows the results of a numerical calculation of the spin waves in magnetite along various symmetry directions using the model parameters in Tables III and IV.

Using the model results, the mode eigenvectors can be analyzed to sort out the spin deviations in each branch. Table V shows the spin deviation amplitude at each atomic site for each mode at $\mathbf{q} = 0$. The acoustic mode ($\omega_1$) has equal amplitude spin precession on each site. The strongly dispersing optic mode has a similar eigenvector to the acoustic mode, but has larger spin deviation on the A site in response to the large internal molecular field. The flat optic branches between 70-80 meV are propagating on the B-sublattice only. Likewise, the flat 130 meV branch propagates on the A-sublattice.

### F. Resolution function convolutions

In order to achieve a complete understanding of the spin wave dispersion and splitting below $T_V$, careful studies of the effects of instrumental resolution must be made. This is especially true considering that the spin wave dispersion is steep, and peakshapes can have long and asymmetric tails due to resolution effects. The resolution must be understood before statements can be made about lifetimes of the excitations obtained from intrinsic peakwidths. Finally, since a fairly accurate model exists for the high temperature spin wave spectrum, we can use the convolutions as a guide to search for phonon excitations that may be overlapping the spin waves.

We begin with the data measured in the cubic phase above the Verwey transition and verify that the Heisenberg model from Ref. 10 agrees with the present data. Using the parameters in Tables III and IV, the acoustic spin wave dispersion along [001] as shown in Fig. 8 is plotted again in Fig. 9(a). Also shown in Fig. 9(a) are the resolution ellipsoids at various (**Q**,ω) points along the dispersion for different instrument configurations. Figures 9(b)-(f) show various measured constant-**Q** energy cuts through the spin waves plotted along with calculations of the resolution convoluted Heisenberg model cross-sections. The general agreement validates the original Heisenberg model.

In order to expedite detailed fits to the spin wave data and extract other parameters such as peakwidths and the position of nearby phonon excitations, an analytical expression for the spin wave dispersion has been developed. The analytical expression is based on a sigmoidal function that reproduces the acoustic spin wave dispersion in the cubic phase of magnetite,

$$\hbar\omega(\mathbf{q}) = Dq_0^2 \left[1 - \frac{q_0^2}{|\mathbf{q}|^2 + q_0^2}\right] \tag{20}$$

Where $D$ is the spin wave stiffness and $q_0$ is the curve shape parameter. We make the assumption (verified by Heisenberg model calculations) that the dispersion is isotropic and has the correct $Dq^2$ limit as $|\mathbf{q}|$ goes to zero. This function is a very good approximation to the dispersion of the acoustic mode as calculated from the Heisenberg model (see Fig. 11). We assume that the spin wave peaks have some intrinsic Lorentzian broadening. Due to the steepness of the dispersion and the relatively broad resolution function, spin waves at many different **q**-values are folded into the convolved cross-section. Therefore, the mode intensities must be well-defined for a good fit. By analysis of the Heisenberg model calculations, the following function reproduces the nearly isotropic spin wave structure factors in the (0,0,4) Brillouin zone,

$$S(\mathbf{Q},\omega) \approx I_0 f^2(Q) \cos\left(\frac{|q|a}{8}\right) \frac{1}{1-\exp(-\hbar\omega/kT)}. \tag{21}$$

Using this function, we are able to fit the high temperature energy scans. During these fits, the parameter $q_0$ was held fixed and the spin wave stiffness, width, and intensity were varied. With $q_0 \approx 0.546$ rlu, typically $D \approx 330$ meV rlu$^{-2}$ and Lorentzian peakwidths of 0.5-2 meV are found. For scans where both spin wave and phonon excitations are present, these were fit simultaneously. Figure 10(a) shows the measured dispersion obtained from fits to all of the data above $T_V$.

The presence of the spin wave splitting below $T_V$ is not captured by the simple Heisenberg Hamiltonian presented in Sec. IIIc. In order to fit the data, we have developed other analytical functions to represent the two (split) branches of the low temperature acoustic spin wave dispersion. The dispersion relations of the split branches branches are labelled U, for the upper branch, and L for the lower branch. When $q < 1/2$, the dispersion relations are fit to the functions

$$\hbar\omega_L(\mathbf{q}) = E_1 \sin^2(qa/2) - E_1' \sin^2(qa) \tag{22a}$$

$$\hbar\omega_U(\mathbf{q}) = E_1 + \Delta. \tag{22b}$$

For $q > 1/2$, the dispersions of the two branches are,

$$\hbar\omega_L(\mathbf{q}) = E_1 \tag{22c}$$

$$\hbar\omega_U(\mathbf{q}) = E_2 - (E_2 - E_1 - \Delta)\sin^2(qa/2) + E_2' \sin^2(qa). \tag{22d}$$

For these functions, $\Delta$ is the energy splitting at $\mathbf{q} = (0,0,1/2)$, giving two modes with energies $E_1$ and $E_1 + \Delta$ at this wavevector. The additional parameters; $E_1'$, $E_2$, and $E_2'$ are used to fit the overall dispersion shape. Typical values of the dispersion parameters determined from fitting are; $E_1 = 40$ meV, $E_1' = 3$ meV, $E_2 = 73$ meV, and $\Delta = 7$ meV.

The two branches $\omega_L$ and $\omega_U$ are plotted in Fig. 11 for typical values of the fitting parameters.

At low temperatures, we used the following functional form for the spin wave mode intensities of branches L and U when $0 < q < 1/2$,

$$S_L(\mathbf{Q},\omega_L) \approx I_1 f^2(Q) \cos\left(\frac{|q|a}{8}\right) \frac{1}{1-\exp(-\hbar\omega_L/kT)} \tag{23a}$$

$$S_U(\mathbf{Q},\omega_U) \approx I_2 f^2(Q) \cos\left(\frac{a}{16}\right) \frac{1}{1-\exp(-\hbar\omega_U/kT)} \tag{23b}$$

for $1/2 < q < 1$,

$$S_L(\mathbf{Q},\omega_L) \approx I_2 f^2(Q) \cos\left(\frac{a}{16}\right) \frac{1}{1-\exp(-\hbar\omega_L/kT)} \tag{23c}$$

$$S_U(\mathbf{Q},\omega_U) \approx I_1 f^2(Q) \cos\left(\frac{|q|a}{8}\right) \frac{1}{1-\exp(-\hbar\omega_U/kT)} \tag{23d}$$

The parameter $I_1$ represents the intensity of the main steeply dispersing and gapped spin wave branch (consisting of the L branch when $q < 1/2$ and the U branch when $q > 1/2$) while $I_2$ is the intensity of any weak and relatively dispersionless extra branch that may arise from symmetry lowering. The intensity functions used in the fitting are associated with the branches as shown Fig. 11.

At $\mathbf{q} = (0,0,1/2)$, the resolution folded fits are shown in Fig. 12 both above and below $T_V$. The rather broad spin wave above $T_V$ becomes two relatively narrow excitations below $T_V$. Based on our resolution convolved fits, the narrowing of the spin waves below $T_V$ is a resolution effect and not a narrowing of the intrinsic width of the peaks. The shape of the low temperature dispersion near $(0,0,1/2)$ flattens out near the gap, no longer being steeply dispersive, giving rise to resolution narrowing effect. The intrinsic width of the spin wave peaks above and below $T_V$ is the same, ~1.5 meV. Fits at other values of **q** and for several experimental configurations are summarized in Fig. 10(b).

Constant-**Q** scans on either side of **q** = (0,0,1/2) show multiple peaks that appear to arise from an additional weak and flatly dispersing excitation branch. Such multiple peaks can be seen in Fig. 4 in the range of **Q** = (0,0,4.35) to (0,0,4.55). However, convolutions of the analytical form in Eqs. (22,23) show that the extra peaks arise from the gapped spin waves at **q** = (0,0,1/2), and not from any additional modes away from (0,0,1/2). In fact, the parameter $I_2$ can be set to zero without seriously affecting the resolution folded peakshapes, as shown in Fig. 12. Thus, if the splitting arises from symmetry lowering or mixing with another excitation, any newly appearing branches are extremely weak. This is indicated in Fig. 10(b), where the dashed horizontal dispersion lines are one possible continuation of the upper and lower spin wave branches.

The combination of steep dispersion, coarse resolution, and fairly large values of neutron incident energy requires detailed knowledge of the full $S(\mathbf{Q},\omega)$ and the resolution function over large regions of $(\mathbf{Q},\omega)$ space. In addition, at the limit of large incident energies, the incident collimation is effectively reduced (and energy resolution improved) because the monochromator is viewed to be smaller at shallow scattering angles. In the coarsest resolution measuremements presented here (configuration D, for example), we are sometimes able to achieve only marginal fits within the Cooper-Nathans formalism. Higher resolution measurements are required to determine the full dispersion of the two branches below $T_V$. In general, the fits above and below $T_V$ reflect the fact that the dispersion is only modified close to **q** = (0,0,1/2). At values of **q** some distance away from (0,0,1/2) the dispersion is the same as that above $T_V$. We also found no substantive difference in the peakwidths above and below $T_V$.

## IV. ANALYSIS OF SPIN WAVES IN MAGNETITE BELOW $T_v$

There are several possible origins of the spin wave splitting in magnetite below the Verwey temperature. A plausible origin of the splitting is due to the lowering of crystallographic symmetry in the Verwey phase. This symmetry lowering causes small changes in the superexchange intergrals due to distortions in the metal-oxygen bond lengths and bond angles in the Fe-O-Fe bonds. If charge ordering is present, then this also modifies the superexchange due to variations in orbital occupancy. In this section, both of these possiblities are considered by developing detailed Heisenberg models in the low symmetry state.

### A. Dependence of the spin wave spectrum on small crystalline distortions

Even in the absence of charge-ordering on the B-sites, the detailed pairwise superexchange values depend on the Fe-O-Fe distances and bond angles and will be modified by the small crystalline distortions. To develop a Heisenberg model in the low symmetry Verwey phase, the orthorhombic space group is used. While the correct space group is likely the monoclinic *Cc* group, it has been shown that all but the three weakest superlattice reflections can be described by the orthorhombic group *Pmca*.[3] The *Pmca* space group does include the (0,0,1/2)-type superlattice reflections (cell doubling) which are of importance in the splitting of the spin wave dispersion. The *Pmca* unit cell of magnetite is indexed relative to the original cubic cell according to the scaling $a/\sqrt{2} \times a/\sqrt{2} \times 2a$. Within this space group, magnetite contains two unique A-sites and four unique B-sites, each with a multiplicity of four giving a total of 24 magnetic iron sites. The collinear moment directions point along the *c*-axis in the *Pmca* space group. Table VI shows the unique Fe and O sites in the *Pcma* structure.

The pairwise exchange values will vary throughout the larger cell due to atomic distortions in the orthorhombic structure. The dependence of the superexchange on bond distances and angles is a difficult theoretical problem, and there is no simple quantitative relationship for its dependence. However, in this instance, we only require knowledge of the small corrections to the superexchange relative to the experimentally determined cubic values. Exact-diagonalization calculations and perturbation theory[29] have shown that for the transition metal oxides, the superexchange is approximately proportional to

$$J_{MN} \propto t_{MO}^2 t_{NO}^2 \cos^2 \theta_{MON} \tag{24}$$

where M,N=A or B and $t_{MO}$ is the metal-oxygen transfer integral. The *pd* transfer integrals depend sensitively on the metal-oxygen distance ($d_{MO}$) as $t_{MO} \propto d_{MO}^{-7/2}$.[30] In the limit of small atomic displacements, we can relate the superexchange of a given M-N pair to the cubic value as

$$\frac{J_{MN}}{J_{MN}^c} \approx \left(1 - \frac{7(\Delta d)_{MO}}{d_{MO}^c} - \frac{7(\Delta d)_{NO}}{d_{NO}^c} - 2(\Delta\theta)_{MON} \tan\theta_{MON}^c \right) \tag{25}$$

where $J_{MN}^c$ is the exchange value in the undistorted cubic structure and $\Delta d$ ($\Delta\theta$) is the bond length (bond angle) deviation from the corresponding cubic value $d^c$ ($\theta^c$). Using this prescription for modifying the superxchange values, we tabulated all of the unique AB exchange paths in the *Pmca* structure. There are 14 distinct Fe-O (AO,BO) pair distances. The AOB bond angles, pair distances, and the corresponding ratio of $J_{AB}/J_{AB}^c$ are given in Table VII. Table VII reveals that some pairs can have their superexchange modified by as much as 30% from the cubic value despite the rather small crystalline distortions from the cubic positions. Using the modified superexchange values, the spin wave dispersion was calculated along the cubic [001] direction and is shown in Fig. 14(a). There are obviously many more branches in the *Pmca* model (24 total), but most of these branches arise from folding the cubic branches into the smaller *Pmca* Brillouin zone. Figure 15 shows that the neutron intensities calculated around the (004) zone from this model show only the two main dispersive cubic branches, all other folded-in branches have very low intensity due to extremely weak structure factors originating from small crystalline distortions. However, the spin wave calculations do show some effects beyond zone folding that depend on the varying exchange values. For example, a reasonably large gap exists in the dispersing optic mode at (0,0,1/2) and the degeneracy of optic modes along the face of the $Fd\bar{3}m$ Brillouin zone are lifted as expected from symmetry lowering. *However, the symmetry lowering model introduces no significant gap in the acoustic spin wave at (0,0,1/2).*

### B. Dependence of the spin wave spectrum on charge ordering

As discussed above, charge order on the B-sites will also influence the pairwise superexchange paths. We must now consider AB' and AB" superexchange paths, where B' is an $Fe^{3+}$ ion (S = 5/2) and B" is an $Fe^{2+}$ ion (S = 2). In order to investigate the effect of charge order on the spin waves, we must know how the $J_{AB'}$ and $J_{AB''}$ superexchange values differ from the average superexchange $J_{AB}$ and we must also know the actual charge ordering pattern. At this point, we ignore the small atomic displacements treated in the previous section and use the cubic atomic positions. The superexchange is [19,31]

$$J_{ij} = \frac{1}{4S_i S_j} \sum_k \frac{2b_k^2}{U_k} \tag{26}$$

for all possible superexchange paths $k$, where $U_k$ is the an effective Coulomb repulsion parameter and $b_k \propto (t^i_{AO} t^j_{BO})_k$. Given that the spinel AOB angle is $125^\circ$, there are hundreds of possible superexchange paths due to the non-orthogonality of the 3$d$-states on the A and B sites. If we assume a $180^\circ$ bond angle, we only need to consider superexchange paths due to $\sigma$-bonds (through the $e_g$ orbitals) and $\pi$-bonds (through the $t_{2g}$ orbitals). This approximation will allow a rough estimation of the dependence of the superexchange on orbital occupancy without detailed analysis using the Slater-Koster integrals and local electron configurational energies. For AB', there is then one $\sigma$-bond and two $\pi$-bonds. For AB", there is again one $\sigma$-bond since the B-site $e_g$ levels are unaffected, but the average number of $\pi$-bonds decreases to 4/3 due to the presence of an extra electron in the octahedral $t_{2g}$ orbital (there are three orbitals that the extra $t_{2g}$ electron can occupy, $xz$, $yz$, $xy$ and a total of four superexchange paths amongst the three possible orbital occupations). Given this simplification, the effect of superexchange on charge ordering is

$$J_{AB'} = -\frac{1}{25}\left(\frac{2b_\sigma^2}{U'} + \frac{4b_\pi^2}{U'}\right), \quad J_{AB''} = -\frac{1}{20}\left(\frac{2b_\sigma^2}{U''} + \frac{4}{3}\frac{2b_\pi^2}{U''}\right). \tag{27}$$

It is estimated that $U' \approx 10$ eV, $U'' \approx 8$ eV and $b_\pi^2 \approx 0.1 b_\sigma^2$,[32,33] we then obtain the following ratios; $2J_{AB'}/(J_{AB'}+J_{AB''}) = 0.90$, $2J_{AB''}/(J_{AB'}+J_{AB''}) = 1.10$. Since the average

AB superexchange is $J_{AB} = -2.4$ meV in the cubic phase, we arrive at $J_{AB'} = -2.1$ meV and $J_{AB''} = -2.7$ meV.

We calculated the spin wave spectrum from several proposed charge-ordering patterns based on the *Cc* space group. In the first case, we examined the 11 possible charge-ordering patterns that are consistent with *Cc* symmetry and also satisfy the Anderson condition (of an average of $2.5^+$ in each tetrahedral cluster of B-sites)[34] and having CO wavevectors of (0,0,1/2). One example of the calculated spin wave dispersion is shown in Fig. 12(b). There are a large number of spin wave branches (96) arising from the *Cc* symmetry of the charge ordering pattern. Similar to the atomic distortion model, most of the branches are related to the cubic branches by Brillouin zone folding. *None of the charge ordering patterns studied that satisfy the Anderson condition introduce any gap in the acoustic wave.*

The second case examined is the pattern obtained from neutron diffraction that has a CO wavevector of (0,0,1) ( called the Wright pattern, after Ref. 5). This charge ordering pattern is also consistent with the *Cc* space group, but does not satisfy the Anderson condition. The Wright pattern creates the spin wave dispersion shown in Fig. 14(c). The charge ordering pattern in the Wright model does introduce a small gap (~1 meV) in the acoustic mode, suggesting that CO with (0,0,1) wavevector is necessary to split the acoustic mode. This makes sense, since folding of the Brillouin zone due to (001) type ordering will create a new Brillouin zone boundary at (0,0,1/2). However, in our model this splitting is very small compared to the observed splitting of 7 meV and cannot fully explain the gap.

We also examined several Heisenberg models where the B-B superexchange was varied according to the different combinations in the charge ordered state; B'-B', B"-B", and B'-B". Even though the BB superexchange is an order of magnitude smaller than the AB superexchange, it was anticipated that the spin wave spectrum would be more sensitive to $J_{BB}$ in the charge-ordered state, since charge ordering occurs on the B sublattice. No such sensitivity was found for the acoustic branch, although models show that the variation of BB superexchange did lift the degeneracy of optical spin waves propagating on the B sublattice (in the range of 70 – 80 meV).

### C. Spin-phonon coupling

We have preliminary evidence that the splitting may be formed from the mixing of the acoustic spin wave with a longitudinal phonon. At high temperatures, we observed a longitudinal optical phonon with energy ~40 meV, as shown in Fig. 16. This phonon branch can be tracked back to the Brillouin zone center with an energy of 43 meV. When the phonon branch crosses the spin wave, there is an enhancement of the phonon structure factor, indicating some mixing (not shown). Below $T_V$, Fig. 17 shows that the spin wave mode at (0,0,1/2) splits, with the lower mode approximately locking in at the energy of the optical phonon. We are in the preliminary stages of the study of this effect, and further experimental work to confirm the mixing of these modes is underway.

## V. DISCUSSION

### A. Relation to other materials

The splitting of the acoustic spin wave branch below $T_V$ is a large effect in magnetite. Similar splittings in the acoustic spin waves have been observed in other systems, such as $UO_2$,[35] $FeF_2$,[36] $La_{1-x}Ca_xMnO_3$,[37] and $YVO_3$.[38] In $UO_2$ and $FeF_2$, splittings of the acoustic spin wave branch are of order 1 meV. These splittings do not appear in concert with a structural phase transition, but are understood to originate from mixing of an acoustic spin wave with a transverse acoustic phonon. In $La_{1-x}Ca_xMnO_3$, many splittings of the acoustic spin wave branch are observed that evolve continuously upon cooling. Such splittings have been attributed to a combination of charge ordering and magnon-phonon coupling. In $YVO_3$, a large (5 meV) splitting of the acoustic spin wave branch is observed after a first-order magnetostructural transition. This transition also causes a large decrease in the spin wave bandwidth. It is proposed that these effects on the spin wave spectrum originate from orbital fluctuations/ordering. The results for $La_{1-x}Ca_xMnO_3$ and $YVO_3$ are similar to the observations discussed here in magnetite. Despite the similarities with magnetite, these other results are discussed in terms of different physical models and the effect of lower symmetry on the spin waves has not been ruled out.

### B. Implications for charge ordering in magnetite

It is a topical question to ask whether charge ordering even exists in magnetite.[7,8] From our results, the observation of a gap in the acoustic spin wave at (0,0,1/2) can be interpreted as originating from CO with a wavevector of (001). However, our best attempts to reproduce the size of the splitting from simple arguments concerning the modification of superexchange due to CO does not predict a large enough gap. Despite our simple estimates of the superexchange variation (by assuming only 180° Fe-O-Fe bond angles), we still feel that we are overestimating the superexchange variation. Other than the (0,0,1/2) splitting, the majority of other spin wave branches are not very different above and below $T_V$ (this is true even of the optical spin wave branches[39]). The larger superexchange variations required to produce a bigger gap would also strongly influence the rest of the spectrum. This is not observed. Thus, it is unlikely that detailed calculations of the superexchange will produce the right size splitting and not affect other spin wave energies away from (0,0,1/2). In other words, the gap appears to be associated with the (0,0,1/2) wavevector, thus more likely originating with the coupling to a phonon or charge-density-wave with a specific wavevector. Furthermore, recent experiments have shown that there is probably not full charge disproportionation on the B-sublattice, rather the valence probably varies from $2.4^+$ to $2.6^+$ site-to-site.[5,6] Thus, the main factor determining the variation of the superxchange due to charge ordering in the ionic model, $(1/S_iS_j)$ in Eqn. (27), is suppressed by covalency. This concurs with neutron diffraction data that see only small variations in the magnetic moment sizes in the Verwey phase.[5,15,40] We are left to the conclusion that such a large spin wave splitting cannot originate from charge ordering in a purely ionic model. Of course, this does not disprove the existence of charge ordering, but rather implies that the splitting has other origins.

In an itinerant electronic model for magnetite, the Verwey phase can be viewed as the formation of a charge-density wave (CDW). Neutron diffraction data[5] and x-ray diffuse scattering data[41] infer that a CDW with wavevector (001) is present in the Verwey phase. Such a nesting wavevector is predicted from bandstructure calculations[42]. As our Heisenberg model studies do indicate that (001)-type ordering will cause a splitting at (0,0,1/2), the CDW mechanism cannot be ruled out. The CDW due to nesting instability will cause an itinerant contribution to $J$, peaking near **q**=(001), and may not affect the rest of the spin wave spectrum to any large degree. More theoretical studies are necessary to determine if the CDW mechanism can explain the results observed here. This being said, it is unlikely that magnetite's magnetic properties should be treated as an itinerant electron system (as opposed to local) since the opening of the electronic gap below $T_V$ [43] does not profoundly affect the bulk magnetic properties or strongly affect the size of $J_{AB}$.

### C. Summary

In summary, we have examined the behavior of the spin wave spectrum of magnetite below the Verwey transition temperature. In the monoclinic phase, the spin waves will be affected by charge-ordering and small crystalline distortions, because both of these modify the superexchange. By studying Heisenberg models with large unit cells containing modified pairwise superexchange values, we have found that some models do produce small gaps (~1 meV) in the acoustic spin wave at (0,0,1/2) (*Pmca* crystallographic distortions and the (001)-type CO pattern), but none reproduce the rather large 7 meV gap observed experimentally. It seems necessary that ordering must have a wavevector of (001) in order to split the acoustic mode at (0,0,1/2). Better estimates of the magnetic superexchange in the low temperature charge-ordered phase are welcome. However, other than the (0,0,1/2) splitting, the majority of other spin wave branches really do not change above and below $T_V$ [39], signifying that the superexchange energy is relatively insensitive to the transition. Thus, it is unlikely that detailed calculations of the superexchange will produce the right size splitting and not affect other spin wave energies away from (0,0,1/2). Other origins of the spin wave splitting are possible, such as charge-density wave formation or a large magnetoelastic coupling (i.e. the mixing of a phonon and spin wave near (0,0,1/2)).

## ACKNOWLEDGMENTS

The authors would like to thank T. Holden, B. Harmon, for useful comments and Zin Tun expert assistance during the Chalk River experiments. Work is supported by the U. S. Department of Energy Office of Science under the following contracts; Ames Laboratory under Contract No. W-7405-Eng-82, Oak Ridge National Laboratory, which is managed by UT-Batelle LLC, under contract No. DE-AC00OR22725 and Brookhaven National Laboratory under contract DE-AC02-98CH10886.

# TABLES

TABLE I. Atomic positions in magnetite at room temperature. The space group is $Fd\bar{3}m$ (No. 227) using the second origin choice ($\bar{4}3m$). The cubic lattice parameter is $a_c = 8.394$ Å.

| Atom | Site | $d_x$ | $d_y$ | $d_z$ |
|------|------|-------|-------|-------|
| A | 8(a), $\bar{4}3m$ | 0 | 0 | 0 |
| B | 16(d), $\bar{3}m$ | 5/8 | 5/8 | 5/8 |
| O | 32(e), $3m$ | 0.379 | 0.379 | 0.379 |

TABLE II. Instruments and configurations used for measurements. The C-5 instrument is located at NRU at Chalk River Laboratories. HB1 and HB3 instruments are located at the High Flux Isotope Reactor at Oak Ridge National Laboratory. Each configuration has a fixed final energy. Collimations are reported as full-width-at-half maximum in minutes of arc. Throughout the paper, configurations will be referred to by the configuration label in the first column.

| Label | Instrument | Magnetic field | $E_f$ (meV) | Monochromator/ analyzer | Horizontal Collimation | Vertical Collimation |
|-------|------------|----------------|-------------|--------------------------|------------------------|----------------------|
| A | C-5 | Horizontal | 14.3 | Be(002)/PG(002) | 36'-30'-48'-120' | 48' |
| B | HB3 | None | 14.78 | PG(002)/PG(002) | 48'-40'-60'-120' | 48' |
| C | HB1 | None | 13.7 | PG(002)/PG(002) | 48'-40'-60'-240' | 180' |
| D | HB1 | None | 13.7 | PG(002)/PG(002) | 48'-80'-80'-240' | 48' |
| E | HB3 | None | 14.87 | Be(002)/PG(002) | 48'-60'-60'-120' | 48' |

TABLE III. Iron positions and spins for magnetite in the primitive cell of the cubic $Fd\bar{3}m$ structure with rhombohedral direct lattice vectors: $(a_c/2, a_c/2, 0)$, $(a_c/2, 0, a_c/2)$, $(0, a_c/2, a_c/2)$.

| i | Fe position | $\sigma_i S_i$ ($\mu_B$) | $d_x$ | $d_y$ | $d_z$ |
|---|---|---|---|---|---|
| 1 | A, $\bar{4}3m$ | -2.5 | 0 | 0 | 0 |
| 2 | A, $\bar{4}3m$ | -2.5 | 1/4 | 1/4 | 1/4 |
| 3 | B, $\bar{3}m$ | 2.25 | -1/8 | -3/8 | -1/8 |
| 4 | B, $\bar{3}m$ | 2.25 | -3/8 | -1/8 | -1/8 |
| 5 | B, $\bar{3}m$ | 2.25 | -1/8 | -1/8 | -3/8 |
| 6 | B, $\bar{3}m$ | 2.25 | -3/8 | -3/8 | -3/8 |

TABLE IV. Nearest-neighbor pairwise superexchange values used for magnetite in the cubic phase (from Ref. 10).

| Pair, (0i;lj) | J(0i:lj) (meV) |
|---|---|
| AB, (01;03) | -2.4 |
| AA, (01;02) | 0.0 |
| BB, (03;04) | 0.24 |

TABLE V. The spin wave eigenvectors at **q** = 0 from the Heisenberg model in the cubic phase of magnetite.

| i | $T_{1i}$ ($\omega_1$=0) | $T_{2i}$ ($\omega_2$=58 meV) | $T_{3i}$ ($\omega_3$=81 meV) | $T_{4i}$ ($\omega_4$=81 meV) | $T_{5i}$ ($\omega_5$=81 meV) | $T_{6i}$ ($\omega_6$=130 meV) |
|---|---|---|---|---|---|---|
| 1 | 0.423 | 0.567 | 0 | 0 | 0 | -0.707 |
| 2 | 0.423 | 0.567 | 0 | 0 | 0 | 0.707 |
| 3 | 0.423 | 0.315 | -0.471 | 0.613 | -0.364 | 0 |
| 4 | 0.423 | 0.315 | -0.471 | -0.613 | 0.364 | 0 |
| 5 | 0.378 | 0.282 | 0.527 | 0.352 | 0.605 | 0 |
| 6 | 0.378 | 0.282 | 0.527 | -0.352 | -0.605 | 0 |

TABLE VI. Iron and oxygen positions in magnetite below the Verwey transition in the *Pmca* space group (from Ref. 12).

| Atom | Site | $d_x$ | $d_y$ | $d_z$ |
|---|---|---|---|---|
| A1 | 4(d), *m* | 0.25 | 0.0049 | 0.0635 |
| A2 | 4(d), *m* | 0.25 | 0.5067 | 0.1887 |
| B1 | 4(b), $\bar{1}$ | 0 | 0.5 | 0 |
| B2 | 4(c), 2 | 0 | 0.0099 | 0.25 |
| B3 | 4(d), *m* | 0.25 | 0.2643 | 0.3789 |
| B4 | 4(d), *m* | 0.25 | 0.7549 | 0.3746 |
| O1 | 4(d), *m* | 0.25 | 0.2630 | -0.0027 |
| O2 | 4(d), *m* | 0.25 | 0.7477 | -0.0009 |
| O3 | 4(d), *m* | 0.25 | 0.2461 | 0.2540 |
| O4 | 4(d), *m* | 0.25 | 0.7696 | 0.2527 |
| O5 | 8(e), 1 | -0.0116 | 0.0089 | 0.1295 |
| O6 | 8(e), 1 | -0.0067 | 0.5050 | 0.1244 |

TABLE VII. Local variations in the the iron-oxygen bond distances and bond angles and the corresponding variation of the superexchange in the *Pmca* structure. The cubic values are $d^c_{AO}$ = 1.876 Å, $d^c_{BO}$ = 2.066 Å, $\theta^c_{AOB}$ = 123.95°.

| $d_{AO}$ (Å) | $d_{BO}$ (Å) | $\theta_{AOB}$ (deg) | $J_{AB}/J_{AB}^c$ |
|---|---|---|---|
| 1.893 | 2.045 | 122.93 | 0.958 |
| 1.908 | 2.025 | 123.58 | 1.003 |
| 1.908 | 2.068 | 121.89 | 0.768 |
| 1.893 | 2.063 | 123.02 | 0.900 |
| 1.908 | 2.078 | 121.58 | 0.719 |
| 1.871 | 2.012 | 127.34 | 1.353 |
| 1.871 | 2.089 | 124.75 | 0.987 |
| 1.894 | 2.049 | 122.12 | 0.895 |
| 1.896 | 2.042 | 122.74 | 0.945 |
| 1.867 | 2.070 | 124.09 | 1.023 |
| 1.867 | 2.032 | 126.82 | 1.298 |
| 1.894 | 2.058 | 123.92 | 0.956 |
| 1.867 | 2.090 | 124.24 | 0.968 |
| 1.896 | 2.100 | 122.39 | 0.730 |

**FIGURE CAPTIONS**

FIG. 1. Scattering plane in horizontal field configuration with a field of 1.1 T is applied along the cubic [001] direction. Black (gray) lines show the Brillouin zone boundaries of the cubic ($T > T_V$) and orthorhombic ($T < T_V$) lattices. Arrows indicate typical wavevectors studied. These wavevectors are equivalent in the cubic phase. In the orthorhombic phase, the cell doubling axis lies along the field and [100]/[001] directions are no longer equivalent.

FIG. 2. Observed scattering from the acoustic spin wave at four cubically-equivalent (0,0,1/2)-type reduced wavevectors (a) (4,0,-1/2), (b) (0,0,4.5), (c) (1/2,0,4) and (d) (4.5,0,0) at $T = 130$ K (cubic, empty circles) and $T = 115$ K (monoclinic, filled circles). Measurements were made in configuration A in a horizontal magnetic field along the cubic [001] direction. The (0,0,1/2) and (1/2,0,0) reduced wavevectors are inequivalent in the Verwey phase. The spin wave excitation is split below $T_V$ at the (0,0,1/2) position (Figs. (a) and (b)), but only weakly split along (1/2,0,0) (Figs. (c) and (d)) demonstrating that splitting occurs primarily along the cell doubling direction. Spin wave intensity is approximately twice as large for wavevectors along [001] (Figs. (b) and (c)) than along [100] (Figs. (a) and (d)) due to the neutron spin cross-section.

FIG. 3. Temperature dependence of (a) the (6,0,1/2) superlattice peak, and (b) the acoustic spin wave intensity at $\mathbf{Q} = (0,0,4.5)$ and $\hbar\omega = 43$ meV measured in configuration

B. The appearance of the superlattice peak below $T_V$ occurs simultaneously with the splitting of the spin wave branch. Slanted arrows indicate cooling/warming.

FIG. 4. (Color) Constant-**Q** energy cuts made for several **Q**-vectors along the [001] direction at (a) 300 K and (b) 115 K, in configuration C. This instrument configuration has no applied magnetic field, thus the spin waves excitations below $T_V$ are averaged over all three orthorhombic directions. The gap is still clear despite this and is indicated by the slanted red line in (b) and the dotted line marks the gap energy at (0,0,4.5) and $T = 115$ K. In each figure, grey circles mark the spin wave peaks. The dashed lines and arrow indicate phonon excitations.

FIG. 5. (Color) Constant energy contour plots of the scattering in the [h0l] plane near the (004) Brillouin zone above $T_V$ ($T = 134$ K, left panels, (a), (b), and (c)) and below $T_V$ ($T = 115$ K, right panels (d), (e), and (f)). Mesh scans performed at constant energies of 39 meV (top row, (a) and (d)), 43 meV (middle row, (b) and (e)), and 46 meV (bottom row, (c) and (f)) were used to construct the contour plots. These energies correspond to the lower peak, gap, and upper peak of the low temperature split spin wave excitation at (0,0,1/2). Measurements were performed in the D configuration with the HB1 spectrometer in zero applied field.

FIG. 6. Constant energy cuts along the (h,0,4.5) direction above $T_V$ ($T = 150$ K, filled circles) and below $T_V$ ($T = 110$ K, empty circles) at (a) $\hbar\omega = 39$ meV, (b) $\hbar\omega = 43$ meV,

and (c) $\hbar\omega$ = 46 meV. Measurements were performed with the B configuration on the HB3 spectrometer in zero applied field.

FIG. 7. Schematic drawing of the local electronic states for an $Fe^{2+}$ ion on the B-site in the cubic phase of magnetite. The free ion term is split by strong cubic crystal field and a relatively weaker trigonal field, resulting in an orbital singlet ground state. This orbital singlet is unsplit by weak spin-orbit interactions, however the doublet excited state is split by this interaction. Subsequent symmetry lowering due to the Verwey transition cannot split the orbital singlet ground state. All states are subjected to Zeeman splitting by the molecular field. Given an approximate value of 350 T, the Zeeman splittings can lead to low lying local excitations of ~100 meV. The lowest dipole excitation of local character that is observable by neutron scattering is ~200 meV. The local excitations are well above the energies considered here.

FIG. 8. Spin wave dispersion of cubic magnetite along the principal symmetry directions as calculated from a Heisenberg model described in the text. Heisenberg parameters were obtained from Ref. 10.

FIG. 9. (a) The acoustic spin wave dispersion of magnetite along [001] and above $T_V$ as calculated from the Heisenberg model of Ref. 10 and projections of the resolution ellipsoids for the various experimental configurations. (b) – (f) show various resolution convoluted cross-sections obtatined from the Heisenberg model as compared to the measured data for the experimental configurations annotated on the figures.

FIG. 10. The acoustic spin wave dispersion of magnetite along [001] as obtained from fits to the resolution folded analytical dispersion relations in various experimental configurations (a) above $T_V$, and (b) below $T_V$. Empty circles are the fitted energies of the spin wave modes. The solid lines are the analytical dispersion relations for typical values of the fitting parameters. The dotted line in (b) indicates a possible continuation of the dispersion of the upper and lower branches.

FIG. 11. (Color) Analytical curves for the acoustic spin wave dispersion along [001] are compared to the Heisenberg model (black). The red curve is calculated from the high temperature formula given in eqn. (20). The blue curves are the two low temperature branches with $\omega_L(\omega_U)$ the lower (upper) branch as calculated in eqn (22). The continuation of these branches is presumed to be flat and is shown by dotted lines. The dispersive parts of $\omega_L$ and $\omega_U$ are primarily cubic modes whose strong intensity is governed by the fitting parameter $I_1$ (solid blue curve) while the weak and flat parts are governed by $I_2$ (dotted blue curve).

FIG. 12. Resolution folded fits to the spin wave excitation at (4,0,-1/2) as measured in experimental configuration A at (a) $T = 130$ K and (b) $T = 115$ K. Open circles are the experimental data and solid lines are resolution folded fits. The fitting gives an intrinsic spin wave width of ~1.5 meV both above and below $T_V$. Thus, the apparent narrowing of the spin wave modes below $T_V$ is a resolution effect due to the flattening of the dispersion near the gap.

FIG. 13. Constant-**Q** energy cuts of the spin wave at **Q**=(0,0,4.4) in configuration C at (a) $T$ = 300 K, and (b) $T$ = 115 K. Dots are the measured data and lines are the result of convolution of the analytical expressions for S(**Q**,ω) with the resolution function. Arrows point to the fitted spin wave energy for each model. While (a) contains only a single excitation, (b) appears to contain three excitations at **Q**=(0,0,4.4). However, the resolution calculations were performed with the parameter $I_2$ = 0 in (b) meaning that only one branch exists in the model at (0,0,4.4). The two higher energy peaks at 38 meV and 44 meV are the split modes at (0,0,4.5) which are picked up by the tail of the resolution function, as shown schematically in the inset. The dispersion of any weak extra branch is difficult to pick up due to the combination of steep dispersion and coarse energy resolution.

FIG. 14. Low-temperature spin wave dispersion for magnetite along the original cubic [001] direction as calculated for a Heisenberg model with superexchange interactions modified by (a) atomic distortions in the *Pmca* structure, (b) charge ordering obeying the Anderson condition and, (c) charge ordering violating the Anderson condition. In each figure, the dashed line corresponds to the original high-temeprature acoustic spin wave branch.

FIG. 15. (Color) Neutron intensities along [001] in the Pmca structure with locally varied superexchange due to atomic distortions. Despite the large number of branches which are folded into the smaller *Pmca* zone, only the main cubic branches show appreciable intensity due to the small atomic distortions in the Verwey phase.

FIG. 16. Transverse and longitudinal constant-**Q** scans through the acoustic spin wave branch at **q** = (0,0,1/2) (parallel to magnetic field direction) and (1/2,0,0) (perpendicular to field) above $T_V$. Measurements are made in configuration A with a magnetic field of 1 *T* applied along the [001] direction. The dashed line is the fit to a spin wave and a phonon excitation using the full triple-axis resolution convolution. The solid line is the spin wave contribution and the dotted line is the phonon contribution to the cross-section. An optical phonon is observed only in the longitudinal scans.

FIG. 17. (a) Transverse and (b) longitudinal scans through the spin wave mode below $T_V$ in configuration A. Convolution fits are shown for the spin wave (solid), phonon (dotted) and total intensity (dashed). The lower branch of the spin wave moves down to the energy of the optical phonon below $T_V$.

**FIGURES**

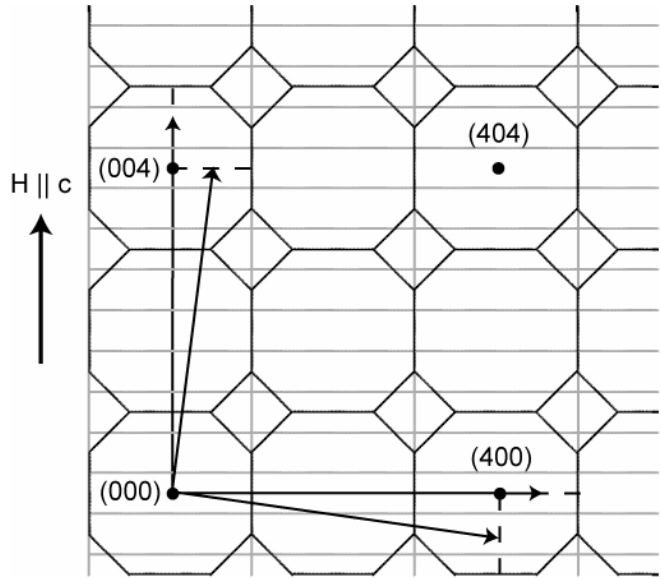

McQueeney, FIG. 1

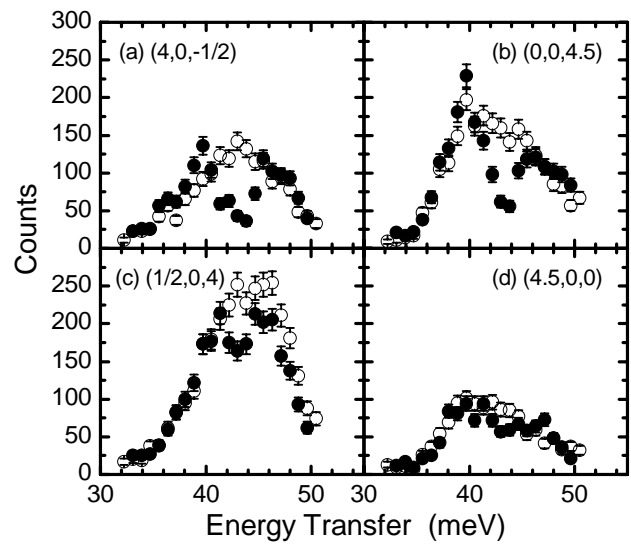

McQueeney, FIG. 2

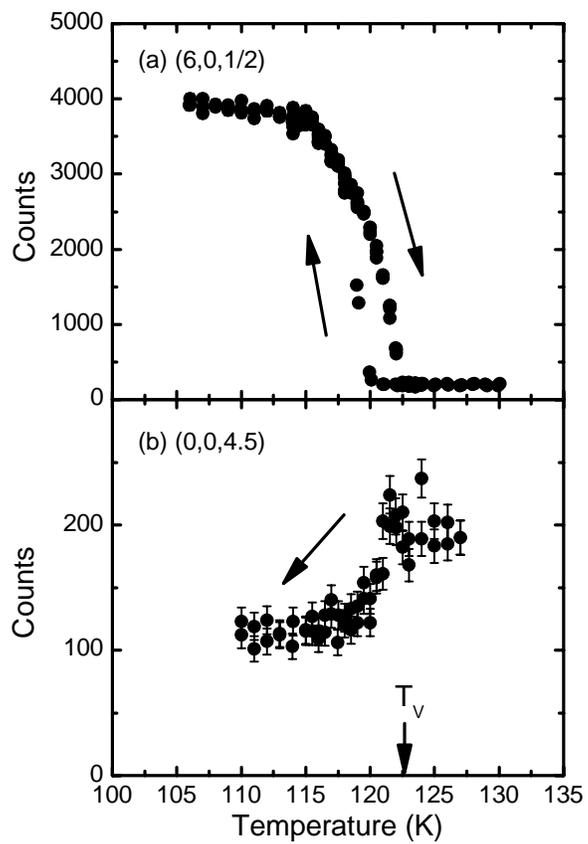

McQueeney, FIG. 3

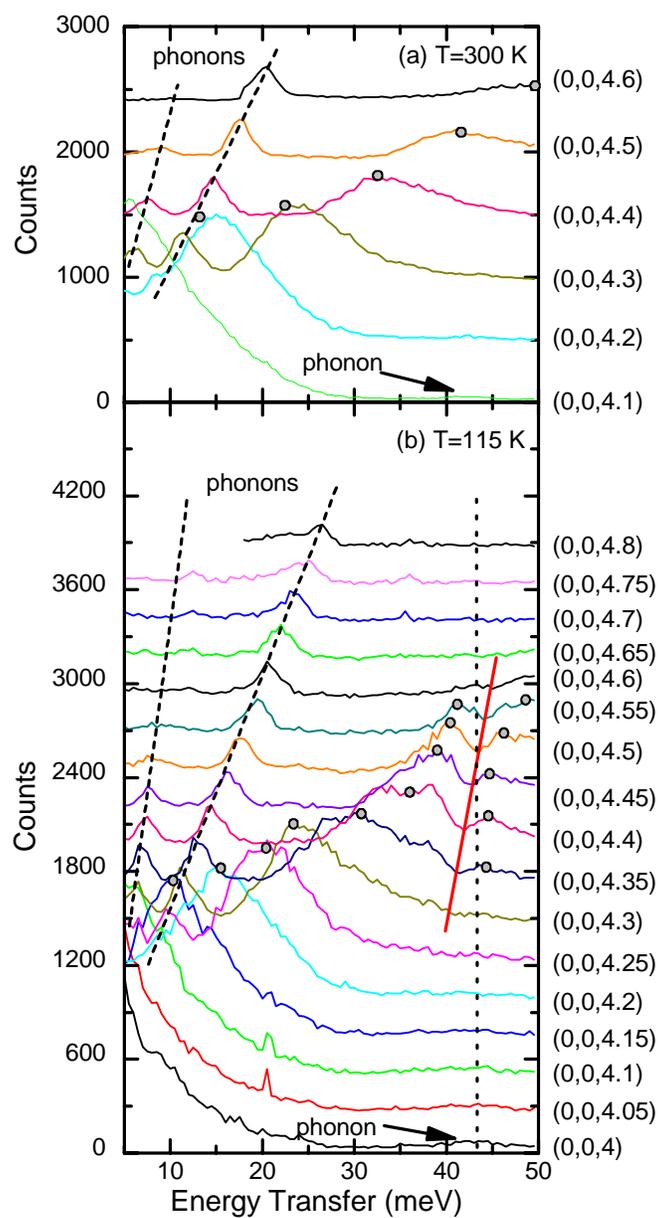

McQueeney, FIG. 4 (Color)

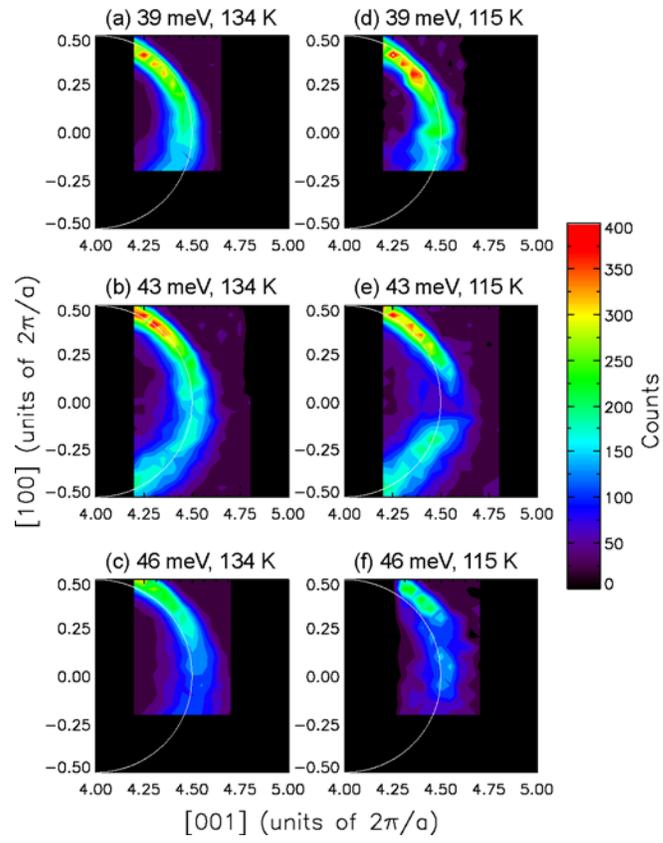

McQueeney, FIG. 5 (Color)

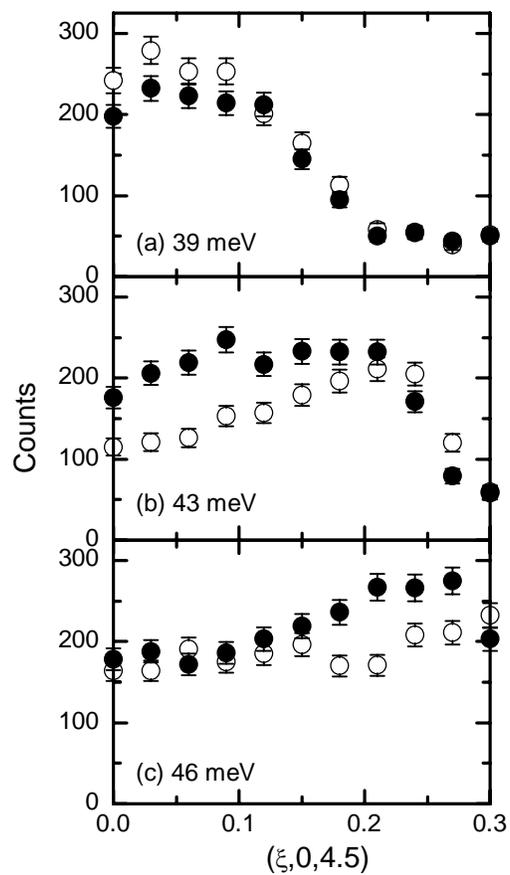

McQueeney, FIG. 6

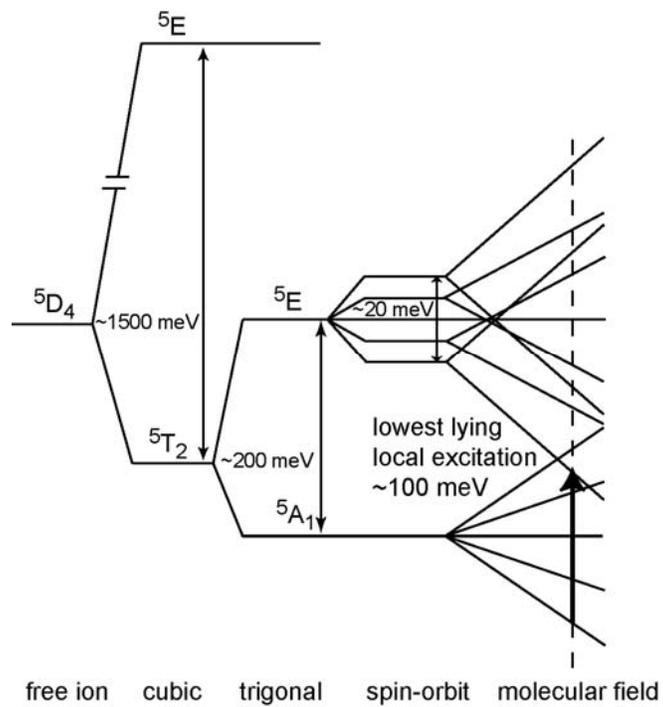

McQueeney, FIG. 7

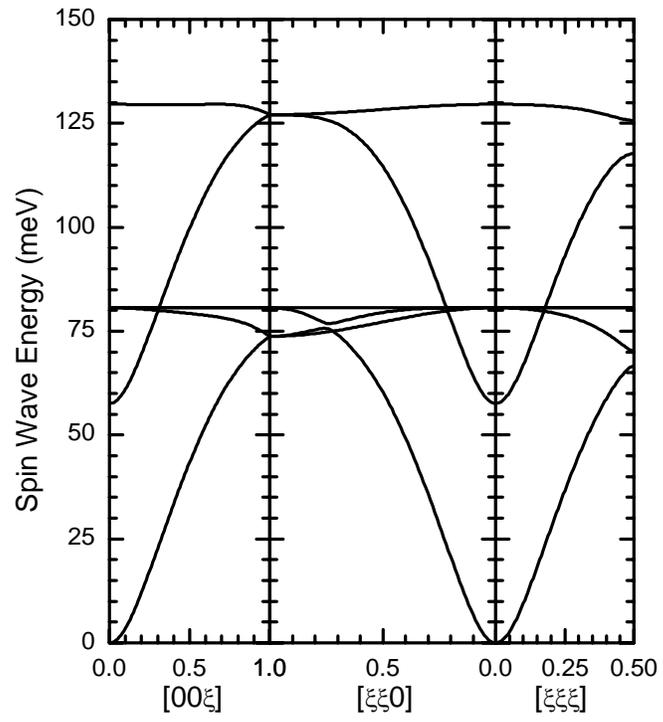

McQueeney, FIG. 8

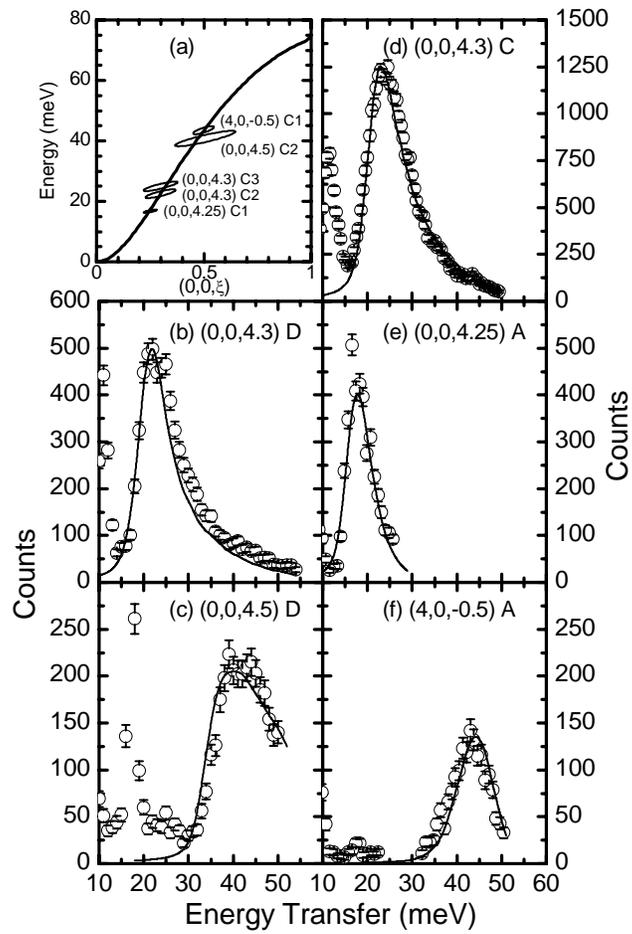

McQueeney, FIG. 9

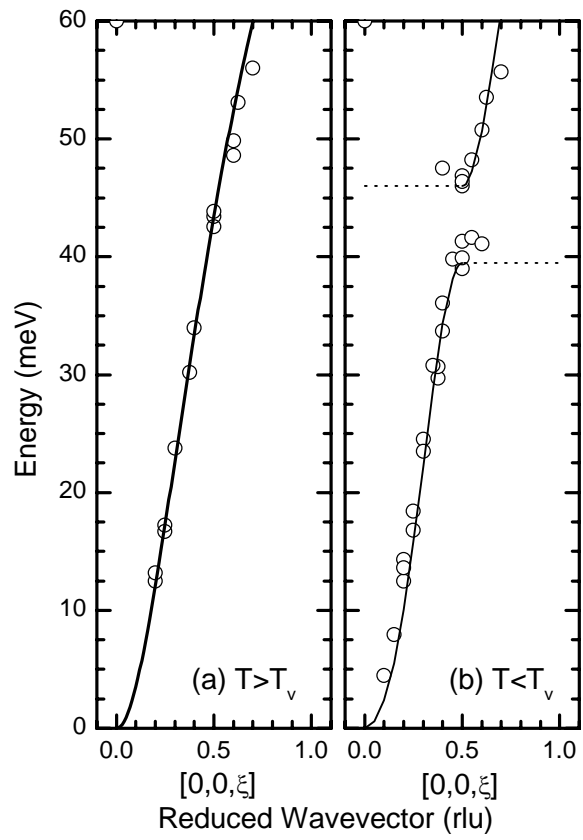

McQueeney, FIG. 10

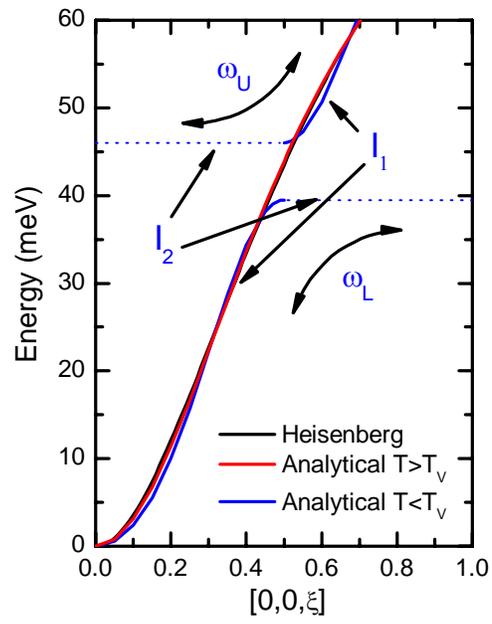

McQueeney, FIG. 11 (Color)

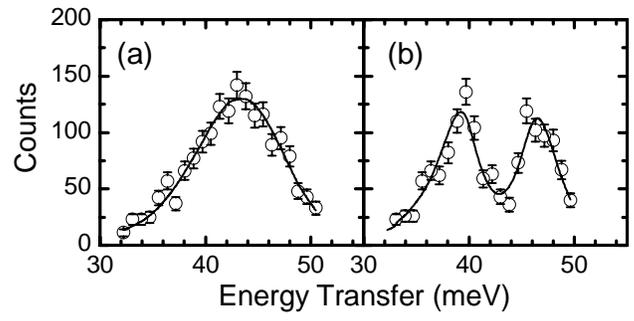

McQueeney, FIG. 12

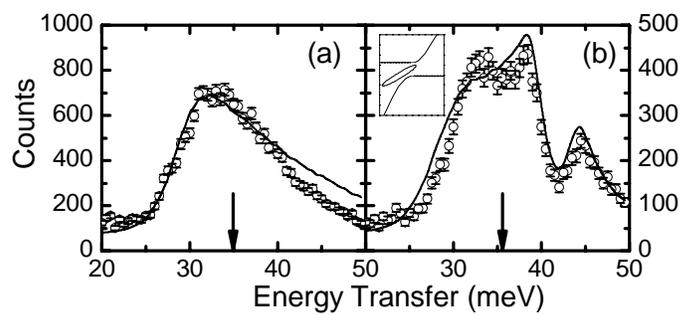

McQueeney, FIG. 13

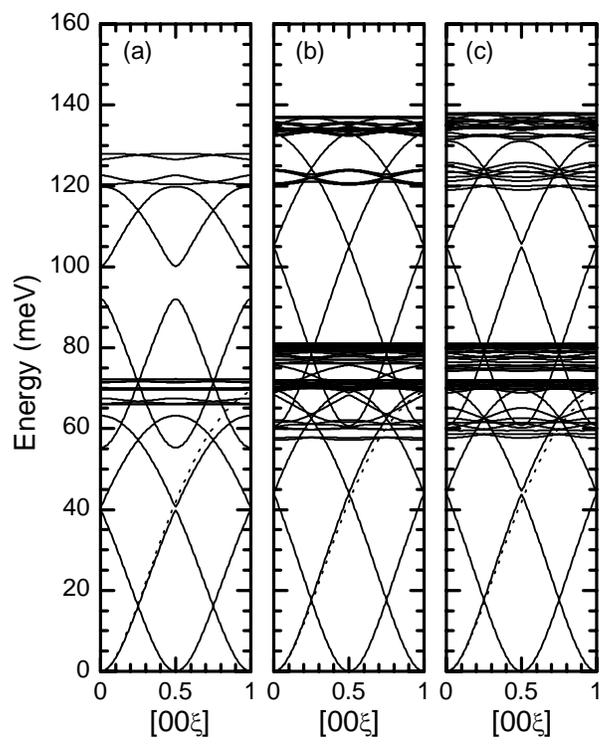

McQueeney, FIG. 14

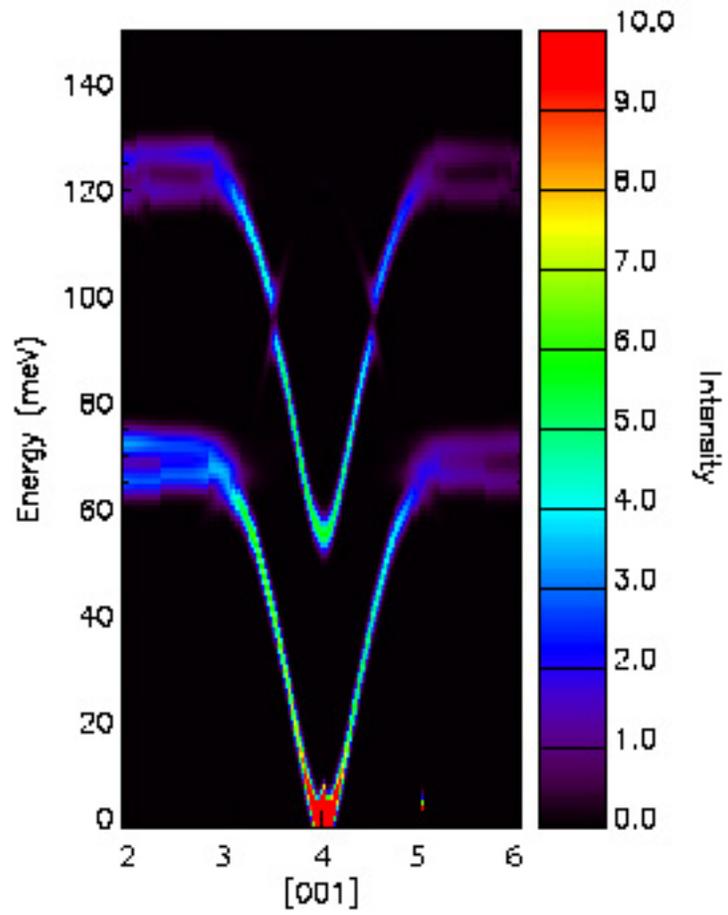

McQueeney, FIG. 15 (Color)

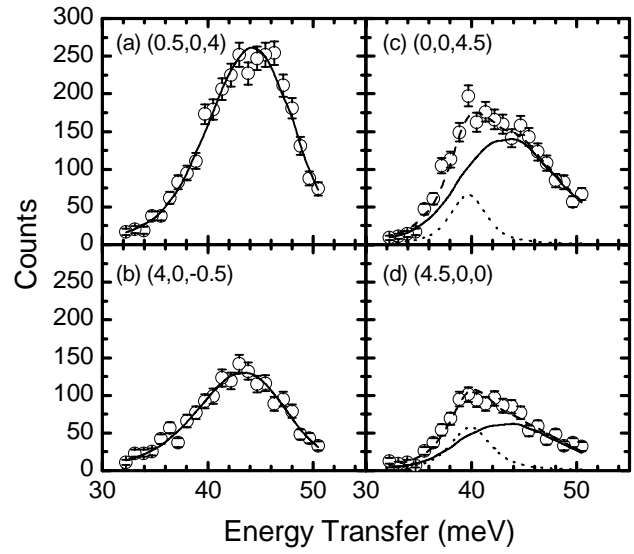

McQueeney, FIG. 16

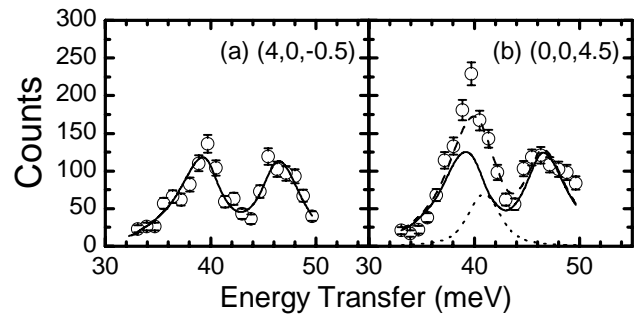

McQueeney, FIG. 17